\documentstyle[prl,twocolumn,aps]{revtex}
\begin{document}
\draft
\twocolumn[
\hsize\textwidth\columnwidth\hsize\csname@twocolumnfalse\endcsname
\preprint{}
\title{Spontaneous flux in a $d$-wave superconductor with 
time-reversal-symmetry-broken pairing state at $\{ 110\}$-oriented 
boundaries} 
\author{Jian-Xin Zhu$^{1}$ and C. S. Ting$^{1,2}$
}
\address{Texas Center for Superconductivity and Department of Physics, 
University of Houston, Houston, Texas 77204\\
$^{2}$National Center for Theoretical Sciences, P.O.Box 2-131, Hsinchu,
Taiwan 300, R.O. China}
%\date{\today}
\maketitle
\begin{abstract}
The induction of an $s$-wave component in a $d$-wave 
superconductor is considered. 
Near the $\{110\}$-oriented edges of such a sample, the induced $s$-wave 
order parameter together with  $d$-wave component forms a complex 
combination $d+e^{i\phi} s$, which 
breaks the time reversal symmetry (BTRS) of the pairing state. As a result, 
the spontaneous current is created. We numerically study the  
current distribution and the formation of the spontaneous flux induced by 
the current. 
We show that the spontaneous flux formed from a number of defect lines 
with $\{110\}$ orientation has a measurable strength. 
This result may provide a unambiguous way to check the existence of BTRS 
pairing state at $\{110\}$-oriented boundaries.
\end{abstract} 
]

\narrowtext
A distinctive feature of a $d$-wave superconductor is its sensitivity  
to inhomogeneities such as impurities, surfaces and 
interfaces. As one of the nodes of the $d$-wave energy gap 
[$\Delta_{d}({\bf k}) \sim \cos k_x a -\cos k_y a$ in the momentum 
space] oriented along $\{110\}$ direction is parallel to the normal 
vector of the surface, Andreev reflected quasiparticles from 
the surface may experience a sign change in the order parameter. 
Quantum constructive interference between the incident and Andreev 
reflected quasiparticle waves then leads to the formation of midgap 
states~\cite{Hu94}, which can explain~\cite{TK95,XMT96} the zero-bias 
conductance peak (ZBCP) observed when tunneling into the 
$\{110\}$-oriented thin films~\cite{GXL88,LGFI92,CSBG96}. Simultaneously, 
while the $d$-wave order parameter is suppressed near the surface, a 
subdominant $s$-wave component of order parameter 
may show up. 
It has been  suggested that a 
broken-time-reversal-symmetry (BTRS) $d+is$ 
pairing state~\cite{Sigrist98} could be generated near a 
surface~\cite{SBL95,Buchholtz95,MS95}
or twin boundary~\cite{SKLM96,FBBK97,BBS98,FY98} with $\{110\}$ orientation. 
While the splitting of the ZBCP observed in a 
copper/insulator/Y-Ba-Cu-O (YBCO) tunneling experiment~\cite{CAPG97} 
would be a signature of BTRS 
pairing state~\cite{MS95,FRS97} formed locally at surfaces, 
other recent tunneling experiments have 
observed no such splitting~\cite{SN98,EDB98,WYGS97}.
Therefore,  the existence of the surface BTRS pairing state in
high-$T_{c}$ superconductors has not been fully established. 
Since it is a weak and secondary effect relative to the ZBCP in the
tunneling experiment, one may argue that the 
samples~\cite{SN98,EDB98,WYGS97} not showing the splitting could be due 
to different surface and measurement conditions. 
Here we suggest to measure the spontaneous 
flux formed near the $\{110\}$-oriented edges as an alternative 
method to confirm the BTRS pairing state.
In this paper, the spontaneous current distribution or flow pattern in a 
square $d$-wave superconductor with $\{110\}$-oriented boundaries will be 
studied.
From this information, we are able to calculate the spontaneous flux 
distribution near the edges. We show that  
the total magnetic flux generated by a bunch of defect lines with 
$\{110\}$ orientation should have a measurable strength. 
Its observation by experiments will unambiguously indicate the existence 
of the BTRS pairing state at the $\{110\}$-oriented boundaries.

We first present a preliminary analysis based upon the Ginzburg-Landau (GL) 
theory to understand how a spontaneous supercurrent is generated by the 
surface BTRS pairing state in a $d$-wave superconductor. 
The GL free energy for a two-component order parameter can be written as  
\begin{eqnarray}
{\cal F}&=&\int d^{2}r \{\alpha_{s}\vert \Delta_{s}\vert^{2} 
+\alpha_{d} \vert \Delta_{d} \vert^{2} +\beta_{1}\vert \Delta_{s}\vert^{4}
+\beta_{2} \vert \Delta_{d}\vert^{4} \nonumber \\
&&+\beta_{3}\vert \Delta_{s}\vert^{2} 
\vert \Delta_{d}\vert^{2} +\beta_{4} (\Delta_{s}^{*2}\Delta_{d}^{2} 
+\Delta_{s}^{2}\Delta_{d}^{*2}) +\gamma_{s} \vert {\bf 
\Pi}\Delta_{s}^{*}\vert^{2} \nonumber \\
&&+ \gamma_{d} \vert {\bf 
\Pi}\Delta_{d}^{*}\vert^{2} 
+\gamma_{v} 
[(\Pi_{x}^{*}\Delta_{s}\Pi_{x}\Delta_{d}^{*}-
\Pi_{y}^{*}\Delta_{s}\Pi_{y}\Delta_{d}^{*})+\mbox{c.c.}]
\nonumber \\
&&+\frac{(\nabla\times{\bf A})^{2}}{8\pi}\}\;,
\end{eqnarray}
where $\Delta_{s,d}^{*}$ are amplitudes of $s$- and $d$-wave components, 
$\alpha_{s,d}$, $\beta_{i}$ ($i=1,2,3,4$), $\gamma_{s,d,v}$ are real 
coefficients, ${\bf A}$ is the vector potential, 
${\bf \Pi}=-i\nabla-2e{\bf A}$ is the gauge invariant momentum operator. 
Note that the free energy here is written in the coordinate system 
defined by the crystalline $a$ and $b$ axes (i.e., $x$ and $y$ axes in 
Fig.~\ref{FIG:CURRENT}) in YBCO. By varying the free energy 
with respect to the vector potential ${\bf A}$, we obtain the current 
\begin{eqnarray}
{\bf J}&=&2e\{ \gamma_{s}\Delta_{s}{\bf \Pi}\Delta_{s}^{*}+
\gamma_{d}\Delta_{d}{\bf \Pi}\Delta_{d}^{*}\nonumber \\ 
&&+\gamma_{v}[\Delta_{s}{\bf \Pi}_{-}
\Delta_{d}^{*}+\Delta_{d}{\bf \Pi}_{-}\Delta_{s}^{*}]\} +\mbox{c.c.}\;,
\end{eqnarray}
where ${\bf \Pi}_{-}=\Pi_{x}{\bf e}_{x}-\Pi_{y}{\bf e}_{y}$ with ${\bf 
e}_{x,y}$ the unit vector along $x$ ($y$) direction. 
We take the normal direction of the surface along the $\tilde{x}$ axis and 
the parallel direction along the $\tilde{y}$ axis. For a $\{100\}$-oriented 
surface where the $\tilde{x},\tilde{y}$ axes coincide with the $a$, $b$ 
crystalline axes, 
the two components of the order parameter are locked into a real 
combination $d\pm s$ by the mixed gradient term in the free energy. In 
this case, no spontaneous current exists. However, for a $\{110\}$-oriented 
surface where $\tilde{x}$ and $\tilde{y}$ are respectively along the 
$y=x$ and $y=-x$ directions as shown in Fig.~\ref{FIG:CURRENT}, 
this mixed gradient term vanishes and an imaginary combination $d\pm is$ is 
required to minimize the free energy. As a result, we find 
\begin{equation}
\label{EQ:CURR-y}
J_{\tilde{y}}=-4e\gamma_{v}\mbox{Im}
(\Delta_{s}\partial_{\tilde{x}}\Delta_{d}^{*}
+\Delta_{d}\partial_{\tilde{x}}\Delta_{s}^{*})\;,
\end{equation}
finite 
[$J_{\tilde{x}}=4e\mbox{Im}(\gamma_{s}
\Delta_{s}\partial_{\tilde{x}}\Delta_{s}^{*}
+\gamma_{d}\Delta_{d} \partial_{\tilde{y}}\Delta_{d}^{*})$ is identically 
zero].  
Away from the surface towards the bulk, the $d$-wave component 
saturates and the $s$-wave component decays to zero.
Thus the parallel spontaneous supercurrent is limited to the  
surface region.  The detailed value of $J_{\tilde{y}}$ must be  obtained by 
solving the GL equation for $\Delta_{s}^{*}$ and 
$\Delta_{d}^{*}$ subject to the appropriate boundary conditions. 
However, the knowledge of $J_{\tilde{y}}$ will not provide the 
information for the flow pattern since the supercurrent must circulate as 
loops in a realistic two dimensional system. 
The investigation of supercurrent flow is a challenging 
topic, it is difficult to employ 
the GL theory or other continuum theories to examine this problem.
The exact diagonalization approach can attack this problem in 
an elegant way. In addition, the band structure and the short 
coherence length effects can also be incorporated. 
The weakness is that the finite size exact diagonalization method, 
if without modification, lacks the spectral resolution 
to resolve the resonances in energy. However, for the order parameter 
and the current (will be calculated below) the finite size effect 
is small since these two physical quantities are determined 
by the local pairing interaction and local hopping 
integral, respectively.   

Within an extended Hubbard model, 
the Bogoliubov-de Gennes (BdG)~\cite{deG66} equations can be written as:
\begin{equation}
\label{EQ:BdG}
\sum_{\bf j} \left( \begin{array}{cc} 
H_{\bf ij} & \Delta_{\bf ij} \\
\Delta_{\bf ij}^{\dagger} & -H_{\bf ij} 
\end{array} \right) \left( \begin{array}{c}
u_{\bf j}^{n} \\ v_{\bf j}^{n} \end{array} \right)
=E_{n} \left( \begin{array}{c}
u_{\bf i}^{n} \\ v_{\bf i}^{n} \end{array} \right)\;,
\end{equation}
where $u_{\bf i}^{n}$ and $v_{\bf i}^{n}$ are the Bogoliubov 
amplitudes at site {\bf i} with eigenvalue $E_{n}$, and 
\begin{equation}
H_{\bf ij}=-t\delta_{{\bf 
i}+\mbox{\boldmath{$\delta$}},{\bf j}}
+(U_{\bf i}-\mu)\delta_{\bf ij}\;,
\end{equation} 
\begin{equation}
\Delta_{\bf ij}=\Delta_{0}({\bf i})\delta_{\bf ij}
+\Delta_{\mbox{\boldmath{$\delta$}}}({\bf i})
\delta_{{\bf i}+\mbox{\boldmath{$\delta$}},{\bf j}}\;.
\end{equation}
Here we choose the $x$ and $y$ 
axes to lie on the diagonals of the sample [as shown in 
Fig.~\ref{FIG:CURRENT}], and 
$\mbox{\boldmath{$\delta$}}=\pm \hat{\bf x}, 
\pm \hat{\bf y}$ are the nearest-neighbor vectors 
along the $x$ and $y$ axes, $t$ is the hopping integral, $U_{\bf i}$ is 
introduced to represent the impurity scattering potential. 
The energy gaps for on-site and nearest-neighbor pairing are determined 
self-consistently     
\begin{equation}
\label{EQ:Gap0}
\Delta_{0}({\bf i})=V_{0}\sum_{n} u_{\bf i}^{n}
v_{\bf i}^{n,*} \tanh (E_{n}/2T)\;, 
\end{equation}
\begin{equation}
\label{EQ:Gap1}
\Delta_{\mbox{\boldmath{$\delta$}}}({\bf i})=\frac{V_{1}}{2}\sum_{n}
[u_{\bf i}^{n}v_{{\bf i}+\mbox{\boldmath{$\delta$}}}^{n,*}
+u_{{\bf i}+\mbox{\boldmath{$\delta$}} }^{n}v_{\bf i}^{n,*}]
\tanh (E_{n}/2T)\;, 
\end{equation}
where $V_{0}$ and $V_{1}$ are 
on-site and nearest-neighbor interaction strength, 
respectively.  
Positive values of $V_{0}$ and $V_{1}$ mean attractive 
interactions and negative values mean repulsive interactions. 
In terms of the Bogoliubov amplitudes, 
the average current from site ${\bf j}$ to ${\bf i}$ is given by 
\begin{equation}
J_{\bf ij}=-\frac{2iet}{\hbar}\sum_{n} \{ 
[f(E_{n})u_{\bf i}^{n*}u_{\bf j}^{n}
+(1-f(E_{n}))v_{\bf i}^{n}v_{\bf j}^{n*}]-\mbox{c.c.}\}\;,
\end{equation}
where the prefactor comes from the spin degeneracy and 
$f(E)=[\exp(E/T)+1]^{-1}$ is the Fermi distribution function. 

We solve the BdG equations ~(\ref{EQ:BdG}) self-consistently~\cite{ZFT99}.
Throughout the work, we take $V_{1}=3t$, 
$\mu=-t$. This set of parameter values give 
$\Delta_{d}=0.368t$, $T_{c}=0.569t$,
and the corresponding coherence length $\xi_{0}=\hbar v_{F}/\pi 
\Delta_{d}\approx 3a$ (In YBCO, $\xi_{0}\approx 15\;\mbox{\AA}$). 
As a model calculation, the value of hopping integral $t$ will be adjusted 
to give a reasonable transition temperature.
The amplitudes of the $d$- and the induced $s$-wave order parameters are 
defined as: 
\begin{mathletters}
\begin{eqnarray}
\Delta_{d}({\bf i})&=&\frac{1}{4}[\Delta_{\hat{\bf x}}({\bf i})
+\Delta_{-\hat{\bf x}}({\bf i})
-\Delta_{\hat{\bf y}}({\bf i})
-\Delta_{-\hat{\bf y}}({\bf i})]\;,  \\
 \Delta_{s}({\bf i})&=&\frac{1}{4}[\Delta_{\hat{\bf x}}({\bf i})+
\Delta_{-\hat{\bf x}}({\bf i})
+\Delta_{\hat{\bf y}}({\bf i})
+\Delta_{-\hat{\bf y}}({\bf i})]\;. 
\end{eqnarray}
\end{mathletters}
In general, both components are complex, i.e., 
$\Delta_{d,s}=\vert \Delta_{d,s} \vert\exp(i\phi_{d,s})$ with 
$\phi_{d,s}\in [0,2\pi]$. 
Their relative phase is defined as $\phi=\phi_{s}-\phi_{d}$.  
We first study the spatial variation of $s$-wave and $d$-wave 
order parameter and the relative phase  
in a square $d$-wave superconductor having $\{110\}$-oriented
boundaries.   The obtained current distribution is drawn in 
Fig.~\ref{FIG:CURRENT}. The current direction is denoted 
by the arrows in the figure. 
This calculation is made under the open boundary condition for 
a clean sample of size $10\sqrt{2}a \times 10\sqrt{2}a$, $T=0.02t$ and 
$V_{0}=-t$. We find that the $d$-wave order parameter is suppressed near 
the boundary and approaches to the bulk value beyond a coherence 
length scale $\xi_{0}$. The induced $s$-wave 
component near the boundary oscillates at an atomic scale and decays
to zero in the bulk region at a distance of a few coherence 
lengths. Near the four corners of the sample, the $d$-wave component is more 
strongly suppressed while the magnitude of induced $s$-wave component is 
enhanced.  Our numerical calculation shows that 
if the order parameter at site ${\bf j}$ within square OHAB (see 
Fig.~\ref{FIG:CURRENT}) is 
$\Delta_{d}=d_r-id_i$ and $\Delta_{s}=s_r+is_i$ 
[for definiteness, $d_{r,i}$ and $s_{r,i}$ are 
taken to be positive], then at site ${\bf j}^{\prime}$ within square 
OHGF which is related to site ${\bf j}$ by a $\pi$/2 rotation around the 
origin O, the order parameter can be obtained 
as $\Delta_{d}=d_r+id_i$ and $\Delta_{s}=-s_r+is_i$. 
Correspondingly, the relative 
phase is $\pi/2+\delta \phi$ at site ${\bf j}$ 
and $\pi/2-\delta \phi$ at site ${\bf j}^{\prime}$, 
where $\delta \phi= \tan^{-1}(s_i/s_r)-\tan^{-1}(d_r/d_i)$ is site 
dependent. On the $\vert x\vert=\vert y \vert$ lines [the dotted lines in 
Fig.~\ref{FIG:CURRENT}], $\delta\phi=0$ and the order parameter becomes 
exactly $d+is$. 
Therefore, the pairing state in a square $d$-wave 
superconductor with $\{110\}$-oriented boundaries is a BTRS $d+e^{i\phi}s$ 
state with $\phi$ varying spatially, 
and the existence of the BTRS state is limited within a region of  
the order of $\xi_{0}$ near the sample edges.  
We have also found that the induced $s$-wave component of 
order parameter decreases with increasing temperature 
and on-site repulsive interaction. For a given on-site interaction 
$V_{0}=-t$, the $s$-component or the surface pairing state vanishes at a 
temperature about $0.1T_{c}$~\cite{ZFT99}.
In our lattice model, the current flows into 
(or out of) each lattice site connecting four bonds. The current 
conservation on each site is respected.  
As it can be seen from 
Fig.~\ref{FIG:CURRENT}, the current flowing at four edges, AC, CE, EG, and
GA, does not form a closed loop to surround the whole sample. Instead, the
current distribution 
has a spatial reflection symmetry with 
respect to the $x$ and $y$ axis. Current flows 
pointing to the corners on the $x$ axis but away from the corners 
on the $y$ axis. At the left and right corners, the current flowing 
in the $x$ direction is separated into two equal branches, which flow in 
the positive and negative $y$ direction; while at the upper and lower 
corners, the current flowing in the $y$ direction consists of two 
equal parts flowing in the positive and negative $x$ direction.  
From an overall point of view, current flows clockwise or 
counterclockwise in four triangles separated by the $x$ and $y$ axes 
[i.e., OAC, OAG, OGE, OEC]. Correspondingly, the total flux in two 
nearest-neighbor triangles for example OAC and OAG has opposite signs. 
In the upper-right inset of Fig.~\ref{FIG:CURRENT}, we indicate the 
direction of flux in each triangle). In contrast to the regular 
diamagnetic Meissner current which flows in a penetration depth,  
the spontaneous current is confined to a distance of a few  
$\xi_{0}$ from the sample edge because it is determined by the spatial 
variation of order parameter. 

To calculate the flux in each triangle, we use the Maxwell 
equation, $\nabla \times {\bf B}=(4\pi/c){\bf J}$, which relates the 
magnetic field ${\bf B}$ to the 
current density ${\bf J}$. 
By assuming that the lattice 
spacing between two consecutive superconducting layers is $c_0$, 
the Maxwell 
equation can be written as 
\begin{equation}
\frac{4\pi}{c}J_{\bf ij}=\frac{c_{0}}{a^{2}}(\Phi_{\bf i}-\Phi_{\bf j})\;,
\label{EQ:MAX}
\end{equation}
where if ${\bf ij}$ is a horizontal link from left to right, $\Phi_{\bf i}$ 
and $\Phi_{\bf j}$ are respectively the magnetic flux passing through 
the upper and lower plaquettes with the link ${\bf ij}$ as a common 
boundary. The magnetic field and the corresponding flux is 
perpendicular to the 
lattice plane since both the current $J_{\bf ij}$ and the vector 
connecting the centers of two plaquettes are in the plane. The flux 
carried by a vortex is then obtained by summing over the flux 
encircled by each plaquette in a triangle. Using 
the calculated current distribution to solve Eq.~(\ref{EQ:MAX}), 
the flux in the triangle AOC is found to be 
$\Phi=\sum_{i}\Phi_{i}= 1.2 {\cal L}I_{0}$, where 
${\cal L}=4\pi a^{2}/cc_{0}$ and $I_{0}=et/\hbar$. For 
$t=0.014\;\mbox{eV}$ which gives $T_{c}\simeq 92.4\;\mbox{K}$, and 
$a=4\;\mbox{\AA}$ and $c_{0}=10\;\mbox{\AA}$, we have  $\Phi=4.04\times
10^{-7} \Phi_{0}$, where $\Phi_{0}=2.07\times 
10^{-7}\;\mbox{G}\cdot\mbox{cm}^{2}$ is the superconducting flux quantum. 
Due to the small size we consider, the value is too small to be 
measurable. 

To measure the spontaneous flux as an evidence for the BTRS pairing 
state, we propose the following experiment. Create several long defect 
lines in a single crystal of $d$-wave superconductor. 
These lines lie parallel to the same $\{110\}$ direction. 
To give a detailed analysis, we consider three parallel 
$\{110\}$-oriented defect lines in the sample of size $100\sqrt{2}a 
\times 100\sqrt{2}a$. The width of them is $\sqrt{2}a$, $\sqrt{2}a$, 
and $3\sqrt{2}a/2$. The centers of the defect lines are located at 
$60a/\sqrt{2}$, $101a/\sqrt{2}$, $159.5a/\sqrt{2}$. Therefore, the 
distance between the edges of two consecutive lines are, 
$39a/\sqrt{2}$, $56a/\sqrt{2}$, respectively. In the numerical 
calculation, we use the single-site potential of strength $U=100t$ to 
model the defect lines, and impose the periodic boundary condition to 
focus on the physical properties associated with the defect lines. The 
spatial dependence of the order parameter and spontaneous current are 
shown in Fig.~\ref{FIG:CURRb}. As expected, when the dominant $d$-wave 
order parameter is suppressed near the defect lines, the $s$-wave order 
parameter is admixed with a relative phase $\pi/2$ with respect to the 
$d$-wave order parameter. More importantly, the admixture of 
the $s$-wave and $d$-wave components is globally $d+is$: 
The relative phase is the same not only on the left and right sides of 
each defect line (i.e., $d+is$:$d+is$) but also near all defect lines.
For the profile of the $s$-admixture near a defect line, a similar result 
that the relative phase is the same on both sides of a $\{110\}$-oriented 
boundary coupling two $d$-wave superconductors was obtained 
based on the quasi-classical theory~\cite{MS95}. 
The same admixture of the $s$-wave component (i.e., $d+is$) on the 
two edges of a superconducting segment created by  
two consecutive defect lines, which comes from a small overlap of 
the electron wavefunctions at two defect lines, can gain more 
condensation energy. If the $s$-wave admixture became $d+is$ at one edge 
and $d-is$ at the other, there would be 
a $\pi$-phase jump in the middle region of the superconducting 
segment, which is physically unacceptable. 
More recently, it has also been found~\cite{HWS99} that an $s$-wave 
component is admixed with the same sign on both sides of a $\{110\}$-oriented 
$d$-wave superconducting strip. This strip is in geometry similar to the 
superconducting segment we are considering here. 
As a consequence, the spontaneous currents flowing at 
the left and right sides of each defect line have opposite directions. 
In this respect, the system under 
consideration is different from an $s$-wave/normal-metal/$d$-wave 
superconductor junction~\cite{HOS97}, where
the BTRS pairing state appears in the normal metal, and the 
spontaneous supercurrent has the same direction 
so that the generated magnetic flux has opposite sign across the 
corresponding region. In addition, our results show that 
the current profile near all defect lines is the same. 
%From the classical electrodynamics, this arrangement of the current 
%directions is also energetically favorable. 
With the above observation, the magnetic field induced by the current 
near these defect lines has the same direction. 
If the pairing state is $d-is$, the directions of the current 
and magnetic field are globally reversed. 
As shown in Fig.~\ref{FIG:CURRb}, our results are 
independent of the width of and the distance between the defect lines. 
Moreover, if the distance between the defect lines is of several 
coherence lengths, the contribution to the magnetic flux from each defect 
line is almost the same. With the chosen values of structure parameter, 
the flux contribution from a single defect 
line of length $30\;\mu m$ is estimated 
to be $3.4\%\Phi_{0}$. If twenty such defect lines are introduced into 
the sample, the total flux can be as large as $68\%\Phi_{0}$. 
In this sense, the effect of BTRS pairing state is greatly amplified 
whereas the information from the tunneling spectroscopy is determined 
by the local property near one individual defect line or surface. 
We suggest to use the heavy-ion 
bombardment technique~\cite{Walter98} 
to realize the desired arrangement of the defect lines. 
With this technique, 
the edge of the defect line can also be very sharp so that it has a well 
defined in-plane orientation. 
The magnetic flux can then be measured by a SQUID microscope 
through a pickup loop~\cite{Kirtley95}. Here the pickup loop 
should be constructed to cover the central part of all defect lines. 
Since the proposed experiment does not involve 
as-grown grain boundaries or twin boundaries, 
the existence of a spontaneous flux would be a 
direct evidence for the BTRS pairing state.   
   
We would like to thank Dr. C. C. Tsuei for useful discussions. 
This work was supported by the Texas Center for Superconductivity 
at the University of Houston, by the Robert A. Welch Foundation, 
and by the grant NSF-INT-9724809.

\begin{figure}
\caption{The current distribution in a square $d$-wave superconductor with  
$\{110\}$ boundaries. The current direction is represented with arrows.
The value of parameters used in the calculation are 
$T=0.02 t$ and $V_{0}=-t$. The upper-right inset shows the direction of
flux in each triangle.
} 
\label{FIG:CURRENT}
\end{figure}

\begin{figure}
\caption{The spatial variation of the order parameter (a) and the 
spontaneous current (b) in a $100\sqrt{2}a\times 100\sqrt{2}a$ sample 
containing three defect lines parallel to the $\{110\}$ direction.  
The solid line (with filled square) corresponds to the $d$-wave component
and the dotted line (with empty square) to the $s$-wave component.
Here the distance is measured in units of $a^{\prime}=a/\sqrt{2}$.
} 
\label{FIG:CURRb}
\end{figure}

\end{document}